\documentclass[conference]{IEEEtran}
\usepackage[a4paper,margin=0.7in]{geometry}
\AtBeginDocument{%
  }

\usepackage{amsmath,amsfonts}
\usepackage{graphicx}
\usepackage{textcomp}
\usepackage{xcolor}
\usepackage{algorithm}
\usepackage{algpseudocode}
\usepackage{listings}
\usepackage{color}
\usepackage{subcaption}
\usepackage{algpseudocode}
\usepackage{multirow}
\usepackage{array}
\usepackage{url}

\date{} 

\begin{document}

%%
%% The "title" command has an optional parameter,
%% allowing the author to define a "short title" to be used in page headers.
\title{\textbf{GPU-Accelerated Simulated Oscillator Ising/Potts Machine Solving Combinatorial Optimization Problems}}

% \author{Yilmaz Ege Gonul, Ceyhun Efe Kayan, Ilknur Mustafazade, Nagarajan Kandasamy, Baris Taskin}
% \affiliation{%
%   \institution{Drexel University}
%   \city{Philadelphia}
%   \state{PA}
%   \country{USA}
% }
% \email{{yeg26, cek99, im445, nk78, bt62}@drexel.edu}

\author{Yilmaz Ege Gonul, Ceyhun Efe Kayan, Ilknur Mustafazade, Nagarajan Kandasamy, Baris Taskin \\\textit{Drexel University, Philadelphia, PA, USA,}\\ Email: \{yeg26, cek99, im445, nk78, bt62\}@drexel.edu}

\maketitle

\footnotetext{
\copyright~2025 ACM. This paper has been accepted for presentation at the Great Lakes Symposium on VLSI (GLSVLSI) 2025. The official version will appear in the GLSVLSI 2025 proceedings.
}

\begin{abstract}
Oscillator-based Ising machines (OIMs) and oscillator-based Potts machines (OPMs) have emerged as promising hardware accelerators for solving NP-hard combinatorial optimization problems by leveraging the phase dynamics of coupled oscillators. In this work, a GPU-accelerated simulated OIM/OPM digital computation framework capable of solving combinatorial optimization problems is presented. The proposed implementation harnesses the parallel processing capabilities of GPUs to simulate large-scale OIM/OPMs, leveraging the advantages of digital computing to offer high precision, programmability, and scalability. The performance of the proposed GPU framework is evaluated on the max-cut problems from the GSET benchmark dataset and graph coloring problems from the SATLIB benchmarks dataset, demonstrating competitive speed and accuracy in tackling large-scale problems. The results from simulations, reaching up to 11295$\times$ speed-up over CPUs with up to 99\% accuracy, establish this framework as a scalable, massively parallelized, and high-fidelity digital realization of OIM/OPMs.
\end{abstract}

\vspace{4pt}


% The preceding line is only needed to identify funding in the first footnote. If that is unneeded, please comment it out.
\section{Introduction}

Combinatorial optimization problems (COPs)~\cite{COP_exponential} represent a critical class of computational challenges that have significant implications across numerous fields, including logistics, telecommunications, bioinformatics, and artificial intelligence. These problems, which involve finding an optimal object from a finite set of objects, are often NP-hard~\cite{COP_exponential}, for which the computational resources required to solve them optimally scale exponentially with problem size. Consequently, there is a persistent need for novel approaches that can provide high-quality solutions within reasonable computational constraints.

\begin{figure*}[t]
   \centering
   \includegraphics[width=\linewidth]{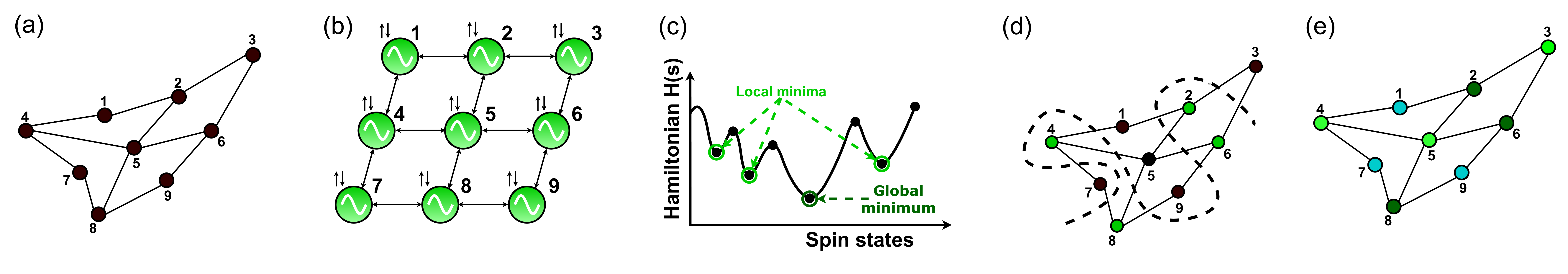}
    \caption{a) An example 9-node graph  b) Problem mapped to a coupled oscillator array c) Energy minimization to ground states d) Max-cut of the graph e) 3-coloring of the graph}
   \label{fig:flow}
\end{figure*}

In recent years, growing interest in physics-inspired computing paradigms that leverage the natural dynamics of physical systems to solve computational problems has led to the emergence of Ising machines~\cite{ising_model}. Ising machines are subject of research as promising specialized solvers for quadratic unconstrained binary optimization (QUBO) problems, which can encode many important combinatorial optimization problems (COPs). Potts machines~\cite{potts_model} are the generalization of the Ising machines, allowing multivalued spins and natively mapping multivalued variable problems. Ising and Potts machines map optimization problems onto systems of coupled spins and utilize the system's tendency to settle into low-energy states to identify optimal or near-optimal solutions.

Various Ising and Potts machine implementations in the literature include quantum annealers~\cite{dwave}, optical parametric oscillator networks~\cite{coherent_potts}, oscillator-based~\cite{wang_oim,seal_potts}, and CMOS implementations~\cite{cmos_annealer,prob_fabric,rtwo_ising, date_potts, iccad_potts}. CMOS-based implementations offer significant advantages, including compatibility with existing semiconductor infrastructure, room-temperature operation, and relatively lower cost of fabrication, along with precision, programmability, and scalability advantages when implemented digitally. Oscillator-based approaches provide natural dynamical systems for optimization with inherent parallelism through phase dynamics that closely match the mathematical properties of Ising and Potts models.
%This paper combines the strengths of both approaches by implementing a numerically simulated, highly scalable oscillator-based Ising/Potts machine that exploits the massive parallel computing capabilities of GPUs. 

In this paper, a GPU-based implementation of a simulated coupled oscillator Potts machine based on the Kuramoto model of coupled oscillators is presented, combining the strengths of CMOS and oscillator-based Ising/Potts machines. The proposed approach leverages CUDA programming to efficiently parallelize the numerical simulation of oscillator dynamics. A modified Kuramoto model~\cite{wang_oim} is employed that incorporates noise-driven stochastic phase transitions and external stimulation for phase stabilization, enabling the system to escape local minima and explore the solution space more effectively. The developed GPU implementation leverages the advantages of digitally computing the oscillator dynamics to propose an Ising/Potts machine framework that achieves floating point precision in representing oscillator phases and coupling weights, scalability in computing on larger graph sizes, and the ability to map graphs of higher densities up to all-to-all connected graphs.

% The remainder of this paper is organized as follows: Section II provides background on the Ising model, the Kuramoto model of coupled oscillators, and their application to combinatorial optimization. Section III details our GPU-accelerated implementation, including the parallelization strategies employed to maximize computational efficiency. Section IV presents experimental results, evaluating the performance of our system on a variety of optimization problems and comparing it with CPU-based implementations. Finally, Section V discusses the implications of our work and potential directions for future research.

\section{Technical Background}
The Ising and Potts models are summarized in Section~\ref{sec:ising_potts}. Mathematical equations that describe the OIM/OPM behavior, utilized in the numerical computations, are shown in Section~\ref{sec:math}.  Mapping of COP problems is presented in Section~\ref{sec:mapping}. The mechanism of SHIL is presented in Section~\ref{sec:SHIL}
The modified Kuramoto model used in the numerical simulations is detailed in Section~\ref{sec:modified_kuramoto}.

\subsection{Ising/Potts Models and Machines}
\label{sec:ising_potts}
\subsubsection{The Ising Model}
The Ising model~\cite{ising_model} consists of discrete variables (spins) that take binary values ($\pm 1$), with interactions between pairs of spins described by coupling coefficients. Mathematically, the energy function (Hamiltonian), excluding the field term, is given by:
\begin{equation}
H_{\text{Ising}} = -\sum_{i<j} J_{ij} s_i s_j 
\end{equation}
where $s_i \in \{-1, +1\}$ represents the state of spin $i$, $J_{ij}$ is the coupling coefficient between spins $i$ and $j$.

\subsubsection{The Potts Model}
The Potts model~\cite{potts_model} generalizes the Ising model by allowing each spin to take one of $q$ possible states ($q \geq 2$), with the Ising model being the special case where $q=2$. The Potts Hamiltonian, excluding the field term, can be written as:
\begin{equation}
H_{\text{Potts}} = -\sum_{i<j} J_{ij} \delta(s_i, s_j)
\end{equation}
where $s_i \in \{1, 2, ..., q\}$ and $\delta$ is the Kronecker delta function.

Many important combinatorial optimization problems, including Max-Cut, Graph Coloring, and Boolean Satisfiability, can be reformulated as Ising or Potts problems~\cite{many_ising}. The goal becomes finding spin configurations that minimize these Hamiltonians, corresponding to optimal solutions of the original problems.

\subsection{OIM and OPM based on the Kuramoto Model of Coupled Oscillators}
\label{sec:math}
Prior study~\cite{wang_oim} mathematically demonstrates that the Kuramoto model of coupled oscillators can be used to represent an Ising/Potts machine through phase interactions of the oscillators. 
In the standard Kuramoto model, the phase evolution of each oscillator in a system of $n$ coupled oscillators is governed by the following differential equation:
\begin{equation}
\frac{d\phi_i}{dt} = \omega_i + \sum_{j=1}^{n} K_{ij} \sin(\phi_i - \phi_j)
\end{equation}
where $\phi_i$ is the phase, and $\omega_i$ is the natural frequency oscillator $i$, and $K_{ij}$ is the coupling strength between oscillators $i$ and $j$. The sine function creates a phase-dependent interaction where oscillators synchronize (when $K_{ij} > 0$) or desynchronize (when $K_{ij} < 0$) depending on the sign of the coupling coefficient.

\subsection{Mapping COPs to Ising/Potts Models}
\label{sec:mapping}
To solve COPs using coupled oscillators, a mapping between the Ising model and the Kuramoto model is proposed in~\cite{wang_oim}. The key insight is that the phases of oscillators can encode the states of the Ising spins, and the coupling coefficients in the Kuramoto model can represent the interaction strengths in the Ising Hamiltonian.

\subsubsection{Max-Cut Problem}
The Max-Cut problem, as shown in Figure~\ref{fig:flow}(d), divides graph vertices into two sets maximizing edge weights on the cut. This maps to the Ising model with $s_i \in \{-1,+1\}$ and $J_{ij} = -1$ for each edge. In oscillator implementations, stable phases $\phi_i \in \{0, \pi\}$ correspond to spin states, with anti-phase synchronization representing the optimal cut.

\subsubsection{Graph Coloring Problem}
Graph Coloring problem, illustrated in Figure~\ref{fig:flow}(e), assigns colors to vertices where no adjacent vertices share colors. The Potts model allows direct mapping~\cite{many_ising} of spin states to colors with $J_{ij} < 0$ for adjacent vertices, making the mapping more efficient than the Ising model, which would require complex penalties and ancillary variables. In oscillator networks, this appears as stable configurations where adjacent oscillators settle into different phases from $\{\frac{2\pi k}{N} | k=0,1,...,N-1\}$, each representing a distinct color.

\subsection{Sub-Harmonic Injection Locking (SHIL)}
\label{sec:SHIL}
Sub-Harmonic Injection Locking (SHIL) is a technique that modifies the standard Kuramoto model by adding a nonlinear drive term to encourage oscillators to settle into specific discrete phase states~\cite{wang_oim}. In the SHIL-enhanced Kuramoto equation:
\begin{equation}
\frac{d\theta_i}{dt} = K \sum_{j=1}^{n} J_{ij} \sin(2\pi(\theta_i - \theta_j)) + K_s \sin(2\pi N \theta_i)
\end{equation}

The additional term $K_s \sin(2\pi N \theta_i)$ creates stable fixed points at evenly spaced phase values. Here, $\theta_i \in [0, 1)$ is the normalized phase of oscillator $i$, $K$ is the coupling strength between oscillators, $K_s$ is the strength of the injection locking term (which can be varied over time), and $N$ determines the number of stable phase states.

This modification enables combinatorial optimization by discretizing the phase space, allowing oscillators to represent discrete variables in both Ising and Potts models. For binary-variable Ising problems ($N=2$), oscillators settle at phases that map to Ising spin states $+1$ and $-1$. For multi-valued Potts problems ($N \geq 3$), oscillators converge to one of $N$ equally spaced phases. As the system evolves according to the coupling matrix $J_{ij}$, the network converges toward minimum-energy configurations as demonstrated in Figure~\ref{fig:flow}(c) that correspond to optimal or near-optimal solutions to the encoded problem.

\subsection{Modified Kuramoto Model for Optimization}
\label{sec:modified_kuramoto}
The proposed GPU-based framework employs a modified version of the Kuramoto model that is specifically tailored~\cite{wang_oim} for solving combinatorial optimization problems with increased accuracy. The dynamics of each oscillator is governed by the following differential equation:

\begin{equation}
\frac{d\phi_i}{dt} = K\sum_{j=1}^{n} J_{ij} \sin(2\pi(\phi_i - \phi_j)) + K_s \sin(2\pi N \phi_i) + \eta_i(t)
\end{equation}

where:
\begin{itemize}
\item $\phi_i \in [0, 1)$ is the normalized phase of oscillator $i$
\item $K$ is the global coupling strength, scaling the strength of every interaction $J_{ij}$
\item $J_{ij}$ is the coupling coefficient between oscillators $i$ and $j$ 
\item $K_s$ is the strength of the influence of the SHIL term
\item $N$ is the number of stable phase states (e.g., $N=2$ for Ising problems)
\item $\eta_i(t)$ is Gaussian white noise with standard deviation $\sigma =\sqrt{K_n \cdot dt}$, where $K_n$ is the noise strength
\end{itemize}

This formulation includes several important modifications to the standard Kuramoto model. Phases are normalized to the range $[0, 1)$ rather than $[0, 2\pi)$ for computational convenience. The SHIL term $K_s \sin(2\pi N \phi_i)$ introduces $N$ stable fixed points at phases $\{\frac{2\pi k}{N} | k=0,1,...,N-1\}$, corresponding to the $N$ possible states of each oscillator phase. Additionally, the addition of Gaussian noise allows the system to escape local minima and explore the solution space more effectively, similar to thermal fluctuations in physical annealing processes.

The Hamiltonian corresponding to this system can be expressed as:
\begin{equation}
H = \sum_{i<j} J_{ij} \cos(2\pi(\phi_i - \phi_j))
\end{equation}
when all oscillators are at stable fixed points (i.e., after thresholding). This Hamiltonian reaches the minimum value when the oscillator phases are configured to minimize the overall energy of the system, which corresponds to the optimal solution of the mapped optimization problem.

\begin{algorithm}
\caption{Simulated Coupled Oscillator Ising/Potts Machine}
\begin{algorithmic}[1]
\Statex \textcolor{blue}{// Initialize variables}
\State $\phi \gets$ Random values in $[0,1)$; $J \gets$ Coupling matrix; $t \gets 0$
\Statex \textcolor{blue}{// Define simulation parameters}
\State $K$, $K_s\_max$, $K_n$ \textcolor{blue}{// Coupling, self-interaction and noise strengths}
\State $h$, $t\_stop$   \textcolor{blue}{// Time step for integration and total simulation time}
\Statex \textcolor{blue}{// Generate triangular waveform for $K_s$}
\State $K_s$ $\gets$ generateTriangularWaveform$(t\_stop, K_s\_max)$
\Statex \textcolor{blue}{// Begin simulation loop}
\While{$t < t\_stop$}
    \Statex \quad \textcolor{blue}{// For each oscillator, compute interaction terms}
    \For{$i = 0$ \textbf{to} $n-1$}
        \State $\partial_i \gets 0$ \textcolor{blue}{// Initialize partial derivative}
        \For{$j = 0$ \textbf{to} $n-1$}
            \State $\partial_i \gets \sum_{j \neq i} J_{ij} \sin(2\pi(\phi_i - \phi_j))$
        \EndFor
    \EndFor
    \Statex \quad \textcolor{blue}{// Update all oscillator phases using Forward Euler method}
    \For{$i = 0$ \textbf{to} $n-1$}
        \State $derivative \gets K \cdot \partial_i + K_s[t] \cdot \sin(2\pi N \phi_i)$
        \State $noise \gets K_n \cdot \mathcal{N}(0,1) \cdot \sqrt{h}$ \textcolor{blue}{// Gaussian noise term}
        \State $\phi_i \gets \phi_i + h \cdot derivative + noise$
            \Statex \quad \quad \textcolor{blue}{// Threshold phases to stable solutions if needed}
        \State $\phi_i \gets$ NormalizeToRange$(\phi_i, 0, 1)$ 
        \textcolor{blue}{// Keep phases in [0,1)}
    \EndFor
    \State $t \gets t + h$  \textcolor{blue}{// Update time}
    
\EndWhile
\end{algorithmic}
\label{alg:sim_coup_osc}
\end{algorithm}

\section{GPU Implementation Details}
The algorithm employed to simulate the coupled oscillator dynamics, CUDA implementation and optimization methods, and the annealing technique used to improve accuracy are detailed in this section.
% The employed algorithm to simulate the coupled oscillator dynamics is detailed in Section~\ref{sec:numerical_sim}. The CUDA implementation and optimization details are presented in Sections~\ref{sec:cuda_implementation} and~\ref{sec:cuda_optimization}. The annealing technique used to escape the local minima is explained in Section~\ref{sec:annealing}.

\subsection{Numerical Simulation Algorithm}
\label{sec:numerical_sim}
The proposed GPU-accelerated framework implements OIM/OPM by efficiently solving the governing differential equations that describe the coupled oscillator dynamics. The implementation utilizes a direct numerical integration approach with the Forward Euler method, inspired by the work in~\cite{emulation_oim} and~\cite{seal_potts}, which offers a balance between computational simplicity and numerical stability. The pseudocode of the algorithm is shown in Algorithm~1. The proposed GPU framework can simultaneously update all oscillator phases in each time step, enabling the simulation of large-scale networks in parallel. 

% The coupled oscillator model offers several advantages. Notably, the model is advantageous for mapping graph problems without preprocessing, as the oscillator array topology can directly mimic the graph structure. Also, the SHIL term allows for straightforward extension to the Potts model by simply adjusting the parameter N, which is difficult to achieve with other modeling techniques due to the difficulty of physically realizing the Kronecker's delta term~\cite{potts_nature}.

The proposed GPU-accelerated algorithm offers several computational advantages. The parallel computation of phase derivatives across all oscillators eliminates the O(n²) sequential bottleneck inherent in CPU implementations. Memory hierarchy optimizations significantly reduce global memory accesses during Hamiltonian calculations. Additionally, the digital implementation enables precise control over the triangular modulation of $K_s$ and exact phase discretization, which are challenging to achieve in analog. The modular structure of the algorithm allows for easy experimentation with different coupling topologies and noise parameters without hardware modifications, enabling rapid exploration of solution spaces for various graphs.

\subsection{CUDA Implementation}
\label{sec:cuda_implementation}
Memory efficiency is achieved by transforming the coupling matrix from 2D to 1D flattened arrays for improved access patterns, storing oscillator phases in separate device memory arrays to prevent race conditions, and utilizing the CURAND \cite{nvidia_curand} library with per-thread generator states for efficient parallel noise generation. Two primary kernels drive the computation: the Kuramoto integration kernel handles the computationally demanding numerical integration of oscillator dynamics by assigning individual oscillators to dedicated GPU threads for computing derivatives, adding noise, and updating phases based on the modified Kuramoto model; while the phase threshold kernel independently processes the phase of each oscillator to find the nearest stable solution. This dual-kernel approach ensures increased parallelism throughout the computation pipeline, enabling efficient simulation of large-scale OIM/OPMs.

\begin{figure}[ht]
    \centering
    \includegraphics[width=0.7\columnwidth]{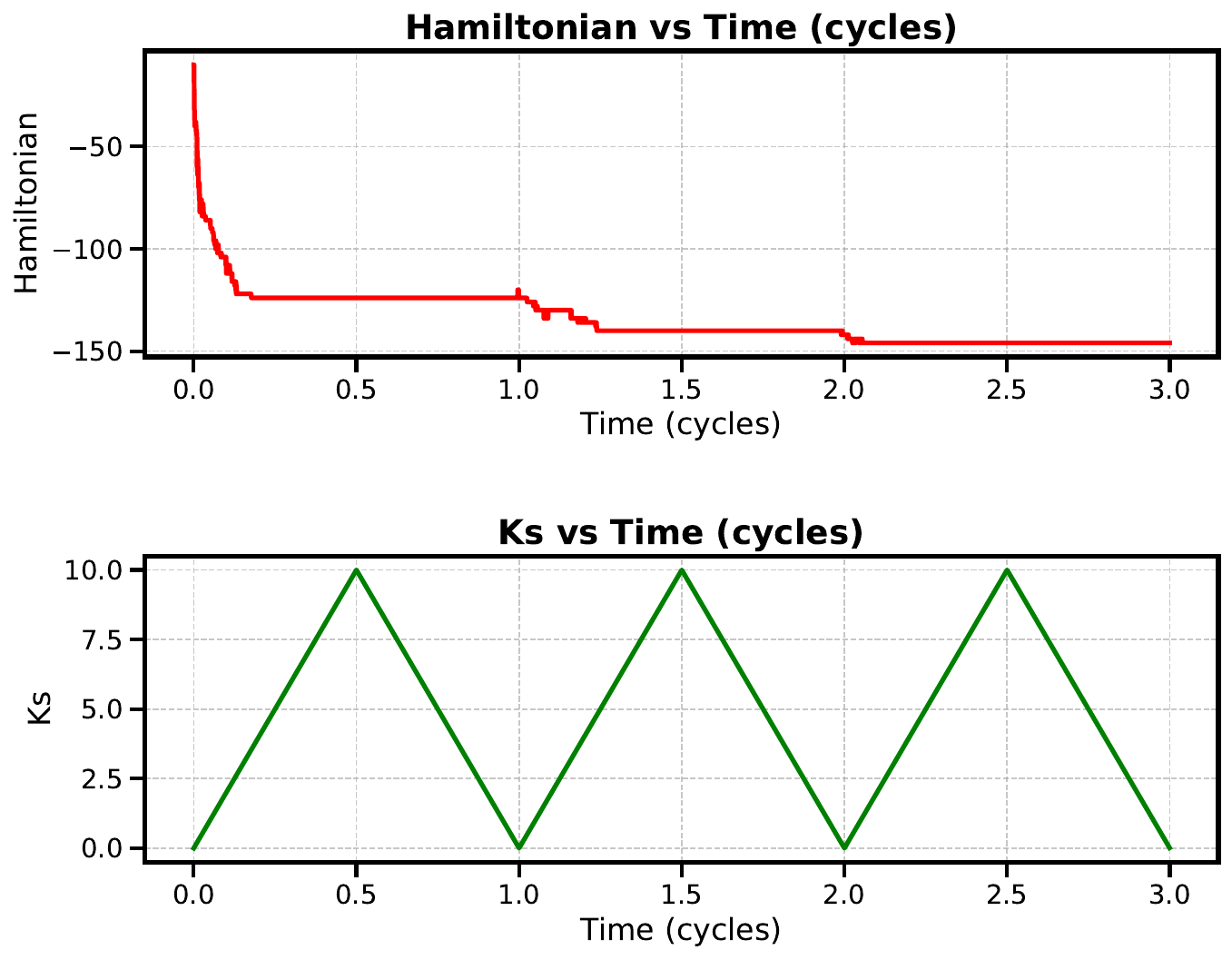}
     \caption{Ks parameter ramped up and down in a triangular waveform and the Ising hamiltonian pushed out of local minima to converge into lower energy states}
    \label{fig:triangle}
\end{figure}

\subsection{CUDA Optimization Techniques}
\label{sec:cuda_optimization}
% The GPU implementation employs parallel derivative computation, shared memory reductions, and memory-efficient buffer swapping to achieve high performance.
The following steps are taken to ensure proper utilization of the memory hierarchy within the GPU: The kernel is tested with float32 instead of double values while maintaining precision, which results in a $\sim$2.46$\times$ improvement in speed compared to the naive version of the implementation.
The largest improvement is observed to come from batch processing of oscillators, rather than a GPU thread per oscillator, resulting in a $\sim$4.6$\times$ improvement. 
Other optimizations include using on-chip shared memory, using cache directives to prioritize L1 usage, unrolling loops to increase register usage, fusing multiplication and addition with CUDA intrinsic instructions, and host-side buffer pinning, which all contribute to another $\sim$2.85$\times$ speedup.

The implementation employs a hierarchical decomposition approach where threads from the same thread blocks cooperatively process batches of oscillator interactions. The batch size can be user-selected and depends on both the GPU and problem size.
The oscillator space is partitioned across multiple blocks.
The block-level batching of the problem allows users to fine-tune their application by selecting the distribution of the batches between thread blocks.

\begin{figure*}[t]
     \centering
     \begin{subfigure}[]{0.78\linewidth}
         \includegraphics[width=\linewidth, height=0.36\linewidth]{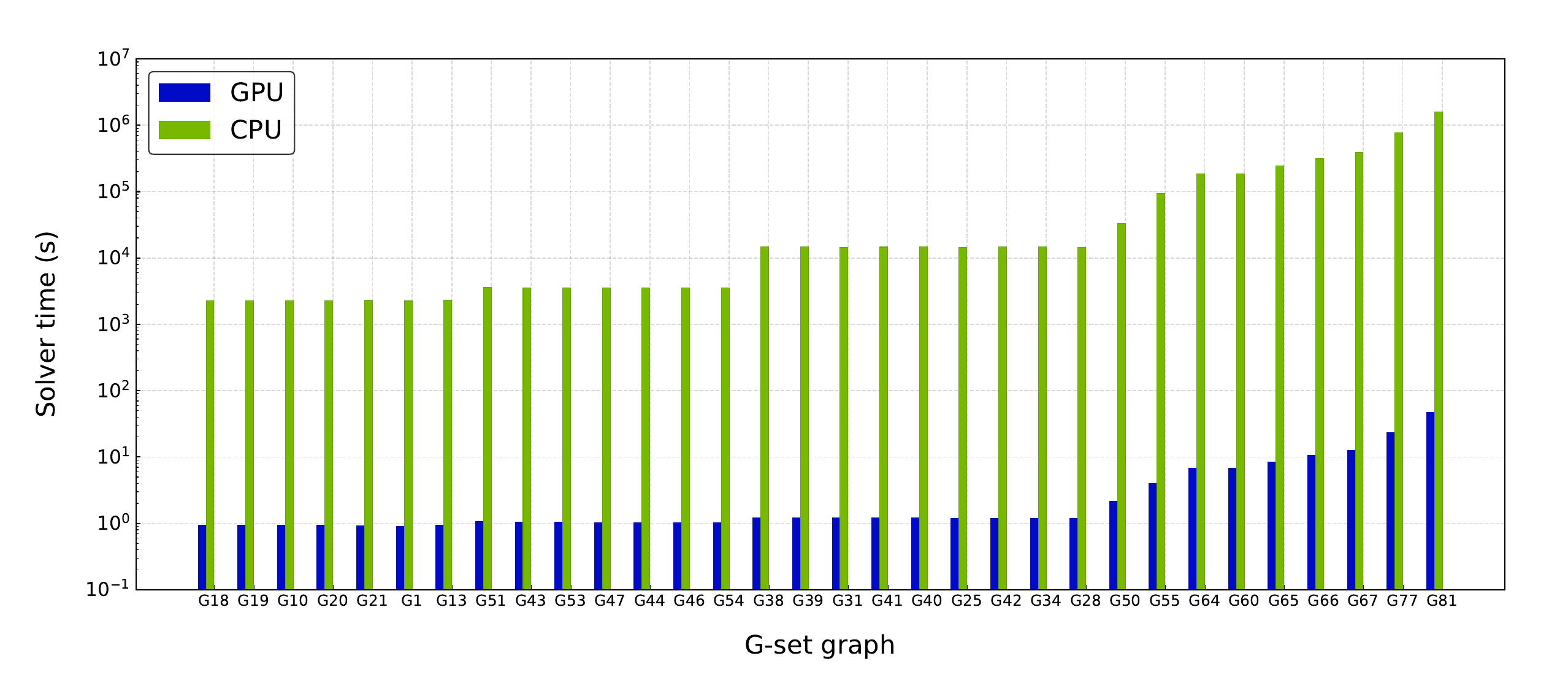}
         \caption{}
    \label{fig:monte_carlo_single}
     \end{subfigure}
     \begin{subfigure}[]{0.4\linewidth}
         \includegraphics[width=\linewidth]{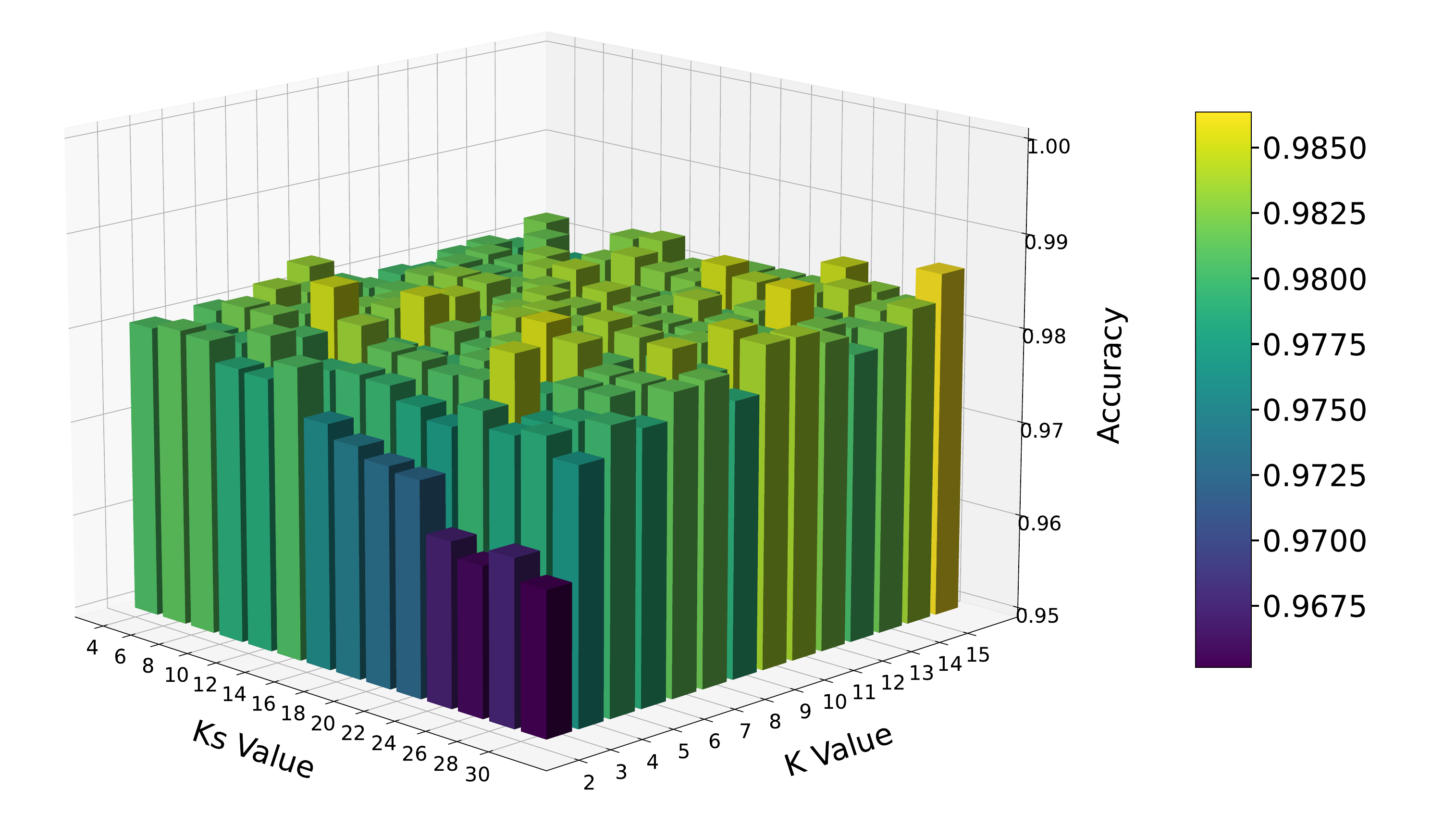}
          \caption{}
         \label{fig:comp_monte_carlo}
     \end{subfigure}
     \begin{subfigure}[]{0.4\linewidth}
         \includegraphics[width=\linewidth]{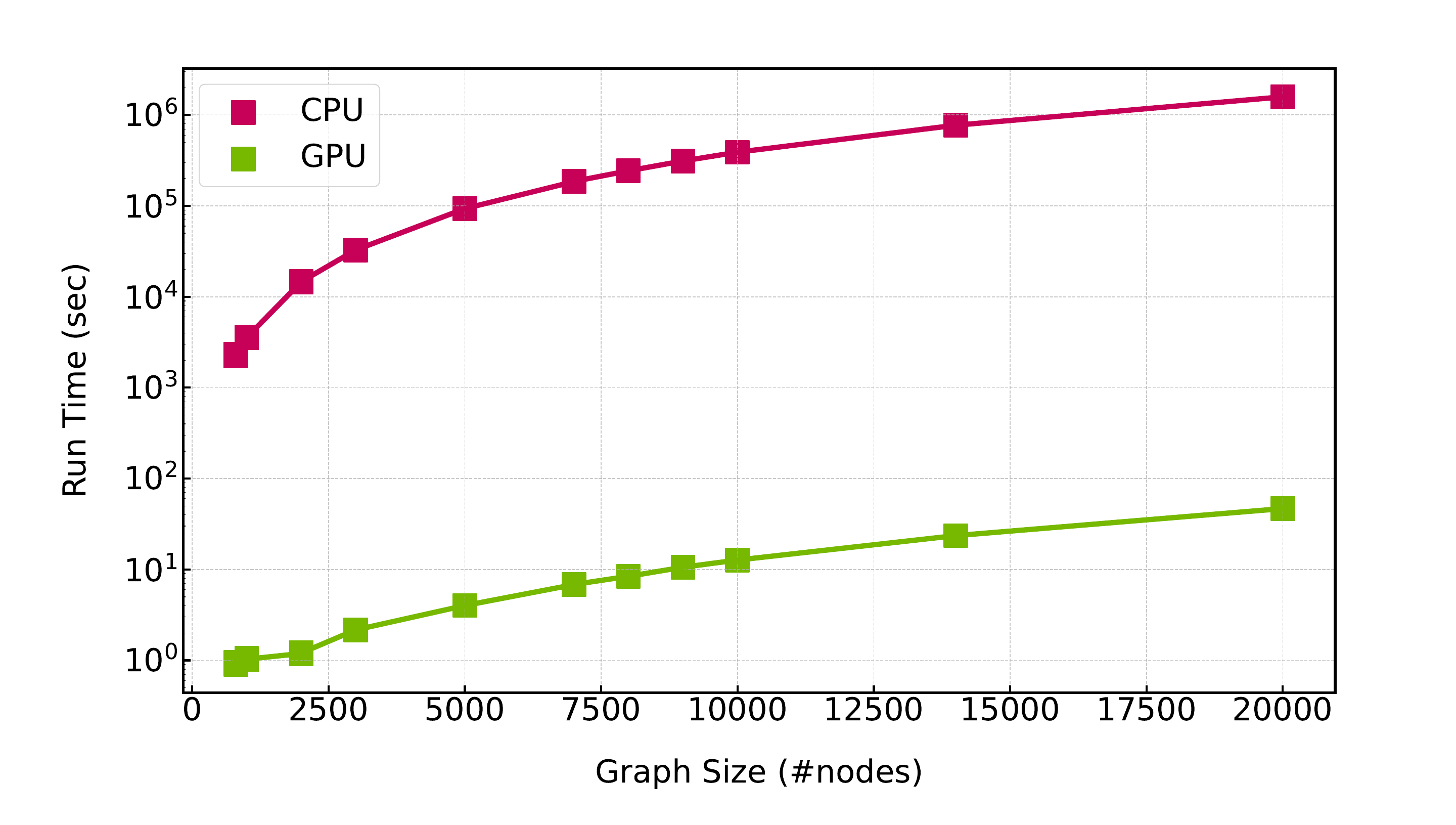}
         \caption{}
         \label{fig:multiple_instances}
     \end{subfigure}
     %  \begin{subfigure}[]{0.48\columnwidth}
     %      \includegraphics[width=\linewidth]{figures/flow_3.pdf}
     %       \caption{}
     % \end{subfigure}
     \caption{a) Speed comparison of CPU vs GPU against the G-set benchmark problems b) Parameter sweep analysis on $K$ and $Ks_{max}$, and accuracy with each combination c) CPU vs GPU run-time scaling in log scale}
    %\label{fig:monte_carlo_single }
    \label{fig:plots}
\end{figure*}

\subsection{Annealing Schedules}

\label{sec:annealing}

The varying Ks parameter in the modified Kuramoto model functions as a critical annealing mechanism that enhances solution accuracy~\cite{wang_oim}. When the Ks value is modulated through a triangular waveform pattern (ramping up and down periodically), the modulation enables the system to thoroughly navigate the solution space rather than becoming prematurely trapped in suboptimal configurations as shown in Figure~\ref{fig:triangle}.
In this work, the annealing schedule is implemented as a triangular waveform for Ks, periodically ramping up to a maximum value and then decreasing to zero.

\section{Experimental Results}

The experimental evaluation of the proposed GPU framework is conducted with simulations on max-cut problems from the G-set benchmark suite~\cite{gset}, and flat 3-coloring problems from the SATLIB benchmark suite~\cite{satlib} and four larger custom-generated 3-colorable graphs. All simulations are executed on a Linux server with an NVIDIA A100 SXM4 GPU with 80GB of memory, leveraging its 6912 CUDA cores to accelerate the parallel computation of oscillator dynamics across large networks. For baseline comparisons and validation, the same algorithm is also implemented in C to run on CPU, executed on the same server with an AMD EPYC 7513 processor featuring 32 cores at 2.6 GHz. The goal of these simulations is to demonstrate the speed of the proposed framework as a digital Ising/Potts solver, while reaching competitive quality of solutions.

It is shown in~\cite{wang_oim} that a coupled oscillator system exhibits an intrinsic convergence time (i.e the number of oscillation cycles to minimize the Hamiltonian) that scales sub-linearly with the oscillator count (i.e. mapped problem size), independent of hardware implementation. In the benchmark results, the stop time of the simulation is adjusted based on the convergence time for the largest problem instance, meaning smaller node graphs could be solved even more efficiently with appropriately calibrated simulation durations.

\subsection{Performance Evaluation}

% Table I presents the performance metrics of our GPU-accelerated implementation across 16 graphs from the G-set benchmark, ranging from 800 to 10,000 nodes. The results demonstrate both the solution quality and computational efficiency of our approach.
% Our implementation consistently achieves high accuracy when compared to the best-known solutions for these benchmark instances. For smaller graphs (up to 3000 nodes), we maintain an accuracy of over 98\%. Even for the largest tested graph with 10,000 nodes, the accuracy remains above 96\%, which is remarkable considering the stochastic nature of the algorithm and the difficulty of these NP-hard problems.

% The runtime performance shows a near-linear scaling with problem size, which is a significant achievement given the O(n²) nature of the all-to-all coupling calculations in the Ising model. For graphs with 1000 nodes, our implementation completes in just over 1 seconds, while the largest graphs with 20,000 nodes require approximately 3 seconds. This demonstrates that our GPU parallelization strategy effectively distributes the computational load.
% An important observation is that the accuracy slightly decreases as the problem size increases. This is expected behavior for heuristic algorithms tackling NP-hard problems and suggests that for very large graphs, additional techniques such as multiple restarts or extended annealing schedules might be beneficial.

Table 1 presents the performance metrics of the GPU-accelerated implementation across 32 max-cut problems (i.e. a 2-spin Ising problem) from the G-set benchmarks, ranging from 800 to 20,000 nodes. 
The proposed framework consistently achieves over 94\% accuracy when compared to the best-known solutions for these benchmark instances, where the worst and best accuracies are 94.08\% and 99.27\% not based on the problem size but based on the problem difficulty. This consistent performance across graph sizes demonstrates the robustness of the framework approach for solving these NP-hard problems, even as the computational complexity increases with larger graphs.

The run-time performance shows near-linear scaling properties with problem size as demonstrated in Figure~\ref{fig:plots}(c). For graphs with 800 nodes, the GPU framework completes in under 0.5 seconds, while the largest graphs with 20,000 nodes require approximately 23 seconds. This represents speed-ups ranging from 795$\times$ for smaller instances to 11,295$\times$ for the largest, as demonstrated in Figure~\ref{fig:plots}(a), showing that the GPU parallelization strategy effectively distributes the computational load across thousands of cores. It is important to note that the accuracy of the GPU and CPU 
implementations are the same as the same algorithm runs on both.

Table 2 presents the performance of the GPU framework on the SATLIB 3-coloring benchmark instances and four custom large 3-colorable (i.e. a 3-spin Potts problem~\cite{iccad_potts}) graphs. Despite the inherently higher complexity of graph-coloring compared to max-cut, with a search space of $3^n$ versus $2^n$ for $n$ nodes where the GPU approach maintains similar accuracy levels with graph-coloring problems. For the standard benchmark instances (flat50-flat200), the GPU achieves speed-ups that increase with problem size, from 1.5$\times$ for the smallest instance to 50$\times$ for the 200-node graph. On the custom larger instances (1000-8000 nodes), the speed-ups grow from 1091$\times$ to 8550$\times$, demonstrating the scaling properties of the GPU implementation for problems of practical size. Execution times remain practical even for the largest 8000-node instance at just 4.72 seconds.
The accuracy remains competitive for the entire test set, with most instances achieving over 98\% accuracy. 

\begin{table}[htbp]
  \centering
  \caption{Accuracy, runtime and speed-up (over CPU) of the GPU framework solving G-set max-cut benchmark problems}
  \resizebox{\columnwidth}{!}{%
  \begin{tabular}{|c||c||c||c||c||c||c|}
    \hline
    Graph & \#Nodes & \begin{tabular}{@{}c@{}}Best\\Known\end{tabular} & GPU-Cut & Acc(\%) & $t_{GPU}$ (s) & SpeedUp \\
    \hline
    \hline
    G81 & 20000 & $<$NA$>$ & 13254 & $<$NA$>$ & 23.37 & 11295x \\
    G77 & 14000 & $<$NA$>$ & 9390  & $<$NA$>$ & 11.79 & 10908x \\
    G67 & 10000 & 6950     & 6546  & 94.19    & 6.34  & 10234x \\
    G66 & 9000  & 6364     & 6018  & 94.56    & 5.29  & 9828x  \\
    G65 & 8000  & 5562     & 5366  & 96.47    & 4.20  & 9640x  \\
    G60 & 7000  & 14188    & 13828 & 97.46    & 3.42  & 9062x  \\
    G64 & 7000  & 8751     & 8313  & 94.99    & 3.40  & 9051x  \\
    G55 & 5000  & 10299    & 10049 & 97.57    & 2.00  & 7756x  \\
    G50 & 3000  & 5880     & 5588  & 95.03    & 1.08  & 5039x  \\
    G28 & 2000  & 3298     & 3149  & 95.48    & 0.58  & 4127x  \\
    G34 & 2000  & 1384     & 1302  & 94.08    & 0.59  & 4125x  \\
    G42 & 2000  & 2481     & 2402  & 96.83    & 0.59  & 4088x  \\
    G25 & 2000  & 13340    & 13173 & 98.75    & 0.59  & 4054x  \\
    G40 & 2000  & 2400     & 2296  & 95.66    & 0.60  & 4053x  \\
    G41 & 2000  & 2405     & 2300  & 95.63    & 0.60  & 4019x  \\
    G31 & 2000  & 3310     & 3125  & 94.41    & 0.60  & 4013x  \\
    G39 & 2000  & 2408     & 2348  & 97.51    & 0.60  & 4011x  \\
    G38 & 2000  & 7688     & 7570  & 98.47    & 0.61  & 4003x  \\
    G54 & 1000  & 3852     & 3793  & 98.47    & 0.51  & 1163x  \\
    G46 & 1000  & 6649     & 6531  & 98.23    & 0.51  & 1158x  \\
    G44 & 1000  & 6650     & 6554  & 98.56    & 0.51  & 1158x  \\
    G47 & 1000  & 6657     & 6566  & 98.63    & 0.52  & 1156x  \\
    G53 & 1000  & 3850     & 3807  & 98.88    & 0.52  & 1144x  \\
    G43 & 1000  & 6660     & 6604  & 99.16    & 0.52  & 1144x  \\
    G51 & 1000  & 3848     & 3789  & 98.47    & 0.53  & 1127x  \\
    G13 & 800   & 582      & 566   & 97.25    & 0.47  & 831x   \\
    G1  & 800   & 11624    & 11539 & 99.27    & 0.45  & 828x   \\
    G21 & 800   & 931      & 880   & 94.52    & 0.47  & 822x   \\
    G20 & 800   & 941      & 900   & 95.64    & 0.47  & 813x   \\
    G10 & 800   & 2000     & 1926  & 96.30    & 0.47  & 806x   \\
    G19 & 800   & 906      & 867   & 95.70    & 0.47  & 805x   \\
    G18 & 800   & 992      & 942   & 94.96    & 0.47  & 795x   \\
    \hline
  \end{tabular}
  }
  \label{tab:graph_metrics}
\end{table}

\begin{table}[h]
    \centering
    \caption{Accuracy, runtime, and speed-up of the SATLIB 3-coloring benchmarks and custom large 3-colorable graphs}
    \resizebox{\columnwidth}{!}{%
    \begin{tabular}{|l||c||c||c||c||c|}
    \hline 
        Graph & \#Nodes & \#Edges & $t_{GPU} (s) $  & SpeedUp & Acc (\%) \\
        \hline
        \hline
        flat50\_115\_1   & 60   & 115   & 0.01 & 1.5x  & 99.1 \\
        flat100\_239\_1  & 100  & 239   & 0.03 & 14x   & 98.7 \\
        flat150\_360\_1  & 150  & 360   & 0.04 & 26x   & 98.2 \\
        flat200\_479\_1  & 200  & 479   & 0.05 & 50x   & 98.8 \\
        custom\_1000     & 1000 & 3453  & 0.53 & 1091x & 96.7 \\
        custom\_2000     & 2000 & 7600  & 0.81 & 3000x & 95.2 \\
        custom\_5000     & 5000 & 21292 & 2.26 & 6857x & 99.2 \\
        custom\_8000     & 8000 & 35948 & 4.72 & 8550x & 99.5 \\
        \hline
    \end{tabular}
    }
    \label{tab:graph_metrics}
\end{table}

\subsection{Parameter Effects on Accuracy and Speed}
Parameter optimization is critical in the numerical simulations. Coupling strength $K$ determines convergence characteristics, where higher values accelerate oscillator information exchange and promote clearer phase separation, improving the solution accuracy while risking numerical instability at excessive levels. SHIL amplitude $Ks$, especially when it is triangle-modulated, functions as an annealing schedule, with time-varying $Ks$ outperforming constant values. The $K/Ks$ ratio must balance the exploration and stability states of the oscillators as shown in Figure~\ref{fig:plots}(b).

Noise amplitude $Kn$ enables escape from local minima when moderate, but degrades solutions when extreme. Time parameters also impact performance, where smaller step sizes $h$ yield more accurate results by reducing truncation errors but increase computational cost, while simulation time $t_{stop}$ must be sufficient for phase synchronization without wasting resources. In the simulations, these parameters are optimized to balance run-time efficiency and solution quality, maximizing $K$ to achieve convergence within a smaller $t_{stop}$.

\begin{table}[t]
    \centering
    \caption{Comparison with related work over benchmark instances}
    \resizebox{\columnwidth}{!}{%
    \begin{tabular}{|c|cc||cc||cc||cc|}
    \hline
    \multirow{2}{*}{} & \multicolumn{2}{c||}{\textbf{G13}} & \multicolumn{2}{c||}{\textbf{G34}} & \multicolumn{2}{c||}{\textbf{G39}} & \multicolumn{2}{c|}{\textbf{G42}} \\
    \cline{2-9}
     & Acc (\%)  & Time (s) & Acc (\%) & Time (s) & Acc (\%) & Time (s) & Acc (\%) & Time (s) \\
    \hline
    \textbf{This work} & 97.25 & 0.47 & 94.08 & 0.59 & 97.51 & 0.60 & 96.83 & 0.59 \\
    \hline
    \textbf{\cite{gpu_ising}} & 90.00 & 0.11 & 86.8 & 0.11 & 94.7 & 0.20 & 94.20 & 0.26 \\
    \hline
    \end{tabular}
    }
    \label{tab:performance_metrics}
\end{table}

\subsection{Comparison with Related Work}
Table 3 presents a comparison between this work and a previous GPU-based Ising solver leveraging a simulated annealing algorithm~\cite{gpu_ising} across four G-set benchmark instances. The proposed framework consistently demonstrates superior solution quality compared to~\cite{gpu_ising}, achieving accuracy improvements of 2.63-7.28 percentage points across all instances, notably due to the capability of the OIM framework to escape from suboptimal solutions. This accuracy advantage comes with a modest increase in computation time (0.47-0.60s versus 0.11-0.26s), that could be higher, considering the use of different GPUs. Nonetheless, the trade-off is favorable for practical applications, as the proposed framework maintains sub-second run times with solutions closer to the global optimum. 

\section{Conclusion}
In this paper, a GPU-based framework for oscillator Ising/Potts machines is presented, solving max-cut and 3-coloring problems with substantial speed-ups over CPU implementations. The proposed GPU-based implementation is a framework for solving combinatorial optimization problems with direct problem mapping, floating point precision coupling, and scalable accommodation of graph size and density. 

 \bibliographystyle{ieeetr}
 \bibliography{ref}

\end{document}